\documentclass[english,aps, prl, twocolumn]{revtex4-1}
\usepackage[T1]{fontenc}
\usepackage[latin9]{inputenc}
\usepackage{amstext}
\usepackage{graphicx}
\usepackage{babel}
\begin{document}

\title{Non-thermal radiation of black holes off canonical typicality}

\author{Yu-Han Ma$^{1,2}$, Qing-Yu Cai$^{3}$, Hui Dong$^{2}$, Chang-Pu
Sun$^{1,2}$}
\email{cpsun@csrc.ac.cn}

\address{$^{1}$Beijing Computational Science Research Center, Beijing 100193,
China~\\
 $^{2}$Graduate School of Chinese Academy of Engineering Physics,
Beijing 100084, China~\\
$^{3}$State Key Laboratory of Magnetic Resonances and Atomic and
Molecular Physics, Wuhan Institute of Physics and Mathematics, Chinese
Academy of Sciences, Wuhan 430071, China}
\begin{abstract}
We study the Hawking radiation of black holes by considering the canonical
typicality. For the universe consisting of black holes and their outer
part, we directly obtain a non-thermal radiation spectrum of an arbitrary
black hole from its entropy, which only depends on a few external
qualities (known as hairs), such as mass, charge, and angular momentum.
Our result shows that the spectrum of the non-thermal radiation is
independent of the detailed quantum tunneling dynamics across the black
hole horizon. We prove that the black hole information paradox is
naturally resolved by taking account the correlation between the black
hole and its radiation in our approach.
\end{abstract}
\maketitle
\textit{Introduction}.-The study of black hole physics have elicited
many important results, such as area law \cite{key-B} and Hawking
radiation \cite{key-sw1,key-sw2}, which have tremendous impacts on
many related researches in different areas of physics \cite{He,Sound-BH,key-Entropyandarea,key-AE}.
One of them is the study of black hole information paradox. The thermal
radiation of black hole directly leads to an information loss \cite{key-sw3,key-Pre},
which contradicts with the properties of unitarity in quantum mechanics. 

In 2000, Parikh and Wilczek \cite{key-PW} considered the problem
of black hole information loss from a consistent perspective of quantum
tunneling, and a non-thermal radiation spectrum was discovered. Such
spectrum allows the correlation between subsequently emitted particles.
Zhang et al. \cite{key-QYC} shows that by taking the correlation
of black hole radiation into account, the black hole information is
conserved. Thus they declared that the black hole information problem
is explained. 

Remarkably, all the radiation spectra studied case by case through
quantum tunneling method \cite{key-18} have a simple form and perfectly
satisfy the requirement of information conservation. This observation
hints an even deep origin of non-thermal nature of Hawing radiation
without referring specially to the geometry of black hole's horizon
\cite{key-HuiDong}, as well as the exact quantum tunneling dynamics.
To reveal such origin of non-thermal spectra, we need an even general
derivation of radiation spectra.

In this letter, we prove that the non-thermal spectrum of Hawking
radiation can be derived with the general principle of canonical typicality
\cite{key-CT1,key-CT2,key-CT3} without referring to the dynamics of
the particle tunneling. The radiation spectrum is directly obtained
by making use of the black hole's entropy and maintaining the non-canonical
part, which matches exactly with the non-thermal feature of the radiation.
Besides black hole, in some specific finite systems \cite{finite system,finite system-1},
we also observed their non-canonical statistic behavior, that is,
their distribution is not a perfect thermal equilibrium distribution.This
implies that the non-thermal property of black hole radiation is the
inevitable result of the finite system statistics that goes slightly off
canonical typicality. With our general formalism, we derive the radiation
spectra of several black holes, which are consistent well with the
previous results achieved from the perspective of quantum tunneling
\cite{key-PW,key-Zhaozheng}. Further, the information carried by
Hawking radiation is discussed in our framework, and then we clarify
the so-called black hole information loss paradox is due to the ignorance
of correlation between the black hole and its radiation. It is worth
mentioning that by introdcing icezones \cite{ice,ice2,ice3} to replace
firewalls \cite{fire1,fire2}, Stojkovic et al. have argued that once
the correlation between the radiation fields is taken into consideration,
the black hole information paradox will no longer exists. Such correlation,
in our approach, is just the primary cause that result in the black
hole radiation off thermal distribution.

\textit{Radiation spectrum off canonical typicality}.-We first consider
the universe U, which consists the system of interest B (e.g. , black
hole) and the environment O (see Fig. 1). The whole universe is assumed
to be in an arbitrary pure universe state

\begin{equation}
\left|\Psi\right\rangle =\sum_{b}\sum_{o}\frac{C\left(b,o\right)}{\sqrt{\Omega_{\textrm{U}}}}\left|b\right\rangle \otimes\left|o\right\rangle ,\label{eq:state}
\end{equation}
where $\Omega_{\textrm{U}}=\Omega_{\textrm{U}}\left(E_{\textrm{U}}\right)$
is the total number of micro-states for the universe with energy $E_{\textrm{U}}$,
and $C\left(b,o\right)$ is the coefficient of state $\left|b\right\rangle \otimes\left|o\right\rangle $.
And $\left|b\right\rangle $ and $\left|o\right\rangle $ are the
eigenstates of B and O, respectively. Without losing generality, we
assume the orthogonal conditions $\left\langle b_{i}|b_{j}\right\rangle =\delta_{ij}$,
and $\left\langle o_{k}|o_{l}\right\rangle =\delta_{kl}$. The normalization
of state $\left|\Psi\right\rangle $, $\sum_{b}\sum_{o}\left|C\left(b,o\right)\right|^{2}/\Omega_{\textrm{U}}=1$,
directly implies the average value of $\left|C\left(b,o\right)\right|^{2}$,
$\overline{\left|C\left(b,o\right)\right|^{2}}=1$ . Let $\left|\Psi_{b}\right\rangle =\sum_{o}C\left(b,o\right)\left|o\right\rangle $.
Then the universe state becomes a maximum entanglement like state,
i.e, $\left|\Psi\right\rangle =\sum_{b}\left|b\right\rangle \left|\Psi_{b}\right\rangle /\sqrt{\Omega_{\textrm{U}}}$.
As a result, ignoring the environment O, the reduced density matrix
of B is obtained as

\begin{equation}
\rho_{\textrm{B}}=\textrm{Tr}_{\textrm{O}}\left(\left|\Psi\right\rangle \left\langle \Psi\right|\right)=\frac{1}{\Omega_{\textrm{U}}}\sum_{b_{i},b_{j}}\left\langle \Psi_{b_{i}}|\Psi_{b_{j}}\right\rangle \left|b_{i}\right\rangle \left\langle b_{j}\right|,\label{eq:rouBa}
\end{equation}
where $\textrm{Tr}_{\textrm{O}}$ means tracing over the variables
of O, and $\left\langle \Psi_{b_{i}}|\Psi_{b_{j}}\right\rangle =\delta_{ij}\left\langle \Psi_{b}|\Psi_{b}\right\rangle $
have been proved in Ref \cite{key-CT2}. For the environment O supported
in a high-dimension Hilbert space, according to central limit theorem,
the average over a large enough subset O of U is as the same as that
over U, so that 

\begin{equation}
\left\langle \Psi_{b}|\Psi_{b}\right\rangle =\sum_{o}\left|C\left(b,o\right)\right|^{2}=\Omega_{\textrm{O}}\left(E_{\textrm{U}}-E_{b}\right).
\end{equation}
Here, $\Omega_{\textrm{O}}\left(E_{\textrm{O}}\right)$ is the number of O's micro-states
with energy $E_{\textrm{O}}=E_{\textrm{U}}-E_{b}$, and $E_{b}$ is the eigenenergy
of $\left|b\right\rangle $. Thus, the reduced density matrix of B
is simplified as

\begin{equation}
\rho_{\textrm{B}}=\sum_{b}\frac{\Omega_{\textrm{O}}\left(E_{\textrm{U}}-E_{b}\right)}{\Omega_{\textrm{U}}}\left|b\right\rangle \left\langle b\right|.
\end{equation}
When the total energy of system B is taken as a certain value, namely, $ E_b=E$, the number of B's micro-states, denoted as  $\Omega_{\textrm{B}}\left(E\right)\equiv\Omega_{\textrm{U}}/\Omega_{\textrm{O}}\left(E_{\textrm{U}}-E\right)$, is fully contributed by the degrees of freedom that degenerated in the macro energy state of B with eigenenergy $E$. When B is specific as a black hole, $E$ corresponds to the total mass $M$ of the black hole. The reduced density matrix of B is thus written as $\rho_{\textrm{B}}=\sum_{b}\left|b\right\rangle \left\langle b\right|/\Omega_{\textrm{B}}\left(E\right),$ which implies that B obeys the micro-canonical distribution. In this
case, the entropy of B, $S_{\textrm{B}}=\ln\Omega_{\textrm{B}}\left(E\right)$,
has been proved to be proportional to the area of B's boundary in
some specific model \cite{key-Entropyandarea, key-AE}. And this is
the so-called entanglement entropy area theorem, which result in a
possible explanation for the origin of the black hole's entropy. 

\begin{center}
\begin{figure}
\includegraphics[width=7.5cm]{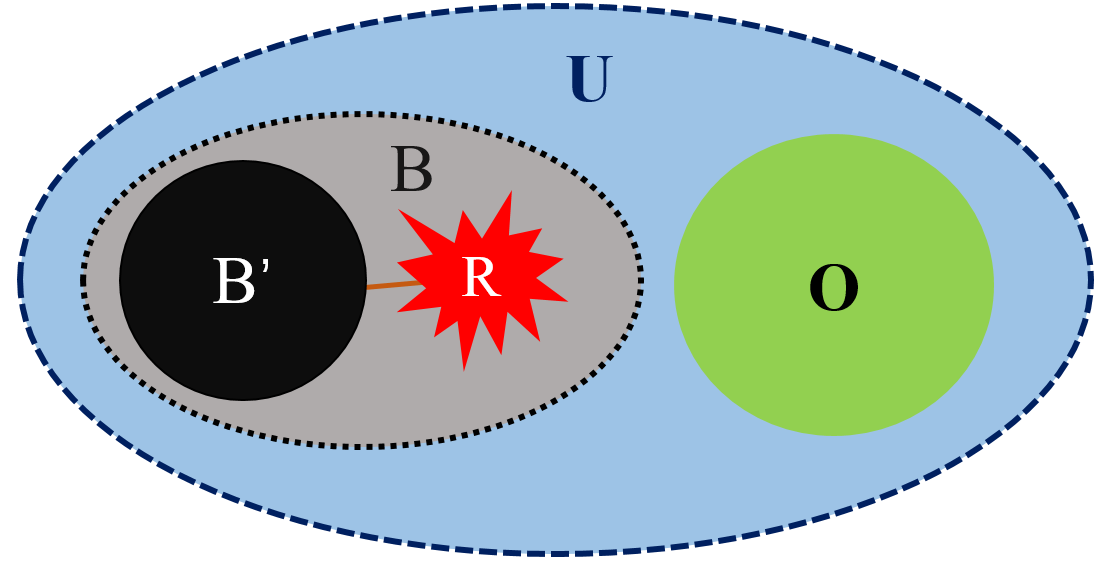}

Figure. 1 (color online). The relation between universe U, the system
of interest B, and the environment O. U consists B and O, and B is
further divided into subsystems R and $\textrm{B}'$
\end{figure}
\par\end{center}

Now we look at B's subsystem R, as shown in Fig. 1, the rest of B
is denoted as $\mathrm{B'=B-R}$. The reduced density matrix of B
can be rewritten as $\sum_{r,b'}\left|r,b'\right\rangle \left\langle r,b'\right|/\Omega_{\textrm{B}}\left(E\right),$
where $\left|r\right\rangle $ and $\left|b'\right\rangle $ are the
eigenstates of R and $\textrm{B}'$ with eigen-energies $E_{r}$ and
$E_{b'}$, respectively. And $E_{r}+E_{b'}=E$ is the constraint condition
given by energy conservation . The reduced density matrix of R

\begin{equation}
\rho_{\textrm{R}}=\textrm{Tr}{}_{\textrm{B}'}\left(\rho_{\textrm{B}}\right)=\sum_{r}\frac{\Omega_{\textrm{B}'}\left(E-E_{r}\right)}{\Omega_{\textrm{B}}\left(E\right)}\left|r\right\rangle \left\langle r\right|,\label{eq:densityR}
\end{equation}
is obtained by tracing over $B'$. Here, $\Omega_{\textrm{B}'}\left(E-E_{r}\right)$
is the number of micro-states of $\textrm{B}'$ with energy $E-E_{r}$,
and it can be rewritten as $\Omega_{\textrm{B}'}\left(E-E_{r}\right)=\exp\left[S_{\textrm{B}'}\left(E-E_{r}\right)\right]$,
where $S_{\textrm{B}'}\left(E-E_{r}\right)$ is the entropy of $\textrm{B}'$
. Therefore, we can further write Eq. (\ref{eq:densityR}) as

\begin{equation}
\rho_{\textrm{R}}=\textrm{Tr}{}_{\textrm{B}'}\left(\rho_{\textrm{B}}\right)=\sum_{r}\textrm{e}^{-\Delta S_{\textrm{BB}'}\left(E_{r},E\right)}\left|r\right\rangle \left\langle r\right|,\label{eq:densityRa}
\end{equation}
where $\Delta S_{\textrm{BB}'}\left(E_{r},E\right)\equiv S_{\textrm{B}}\left(E\right)-S_{\textrm{B}'}\left(E-E_{r}\right)$
is the difference in entropy between B and $\textrm{B}'$ . We can
clearly see from Eq. (\ref{eq:densityRa}) that only when $\Delta S_{\textrm{BB}'}$
depends on $E_{r}$ linearly , the spectrum of R is perfectly thermal.
What should be mentioned here is that we do not expand $\Delta S_{\textrm{BB}'}$
only up to the first order of $E_{r}$, as done in most studies in
the thermodynamic limit. It will be shown later that the higher order
of $E_{r}$ or the non-canonical part of $\rho_{\textrm{R}}$ just
determines the non-thermal property of R.

Until here, the discussion is a general one without any specification
of black hole radiation. Now we will specify our system with B as
black hole, R as Hawking radiation, and $\textrm{B}'$, in this case,
is considered as the remaining black hole. For the black hole B with
three ``hairs'', mass $M$, change $Q$, and angular momentum $J$,
we let $\left|\omega,q,j\right\rangle $ being the eigenstate of radiation
R with energy $\omega$, change $q$, and angular momentum $j$. Then
it follows from Eq. (\ref{eq:densityRa}) and the conservation laws
for charge and angular momentum that the radiation spectrum is obtained
as

\begin{equation}
\rho_{\textrm{R}}=\sum_{\omega,q,j}\textrm{e}^{-\Delta S_{\textrm{BB}'}\left(\omega,q,j,M,Q,J\right)}\left|\omega,q,j\right\rangle \left\langle \omega,q,j\right|.\label{eq:d-1}
\end{equation}
This is the main result of this letter. To get the explicit expression
of $\rho_{\textrm{R}}$, we then make use of the Bekenstein-Hawking
entropy, which reads

\begin{equation}
S_{BH}\left(M,Q,J\right)=\frac{A_{H}}{4}=\pi R_{H}^{2},\label{eq:BH entropy-1}
\end{equation}
where $A_{H}=A_{H}\left(M,Q,J\right)$ and $R_{H}=R_{H}\left(M,Q,J\right)$
are the ``hairs''-determined area and radius of the black hole's
event horizon, respectively. It follows from Eqs. (\ref{eq:d-1})
and (\ref{eq:BH entropy-1}) that the radiation spectrum of the black
hole B is
\begin{widetext}
\begin{equation}
\rho_{\textrm{R}}=\sum_{\omega,q,j}\exp\left[\pi R_{H}^{2}\left(M-\omega,Q-q,J-j\right)-\pi R_{H}^{2}\left(M,Q,J\right)\right]\left|\omega,q,j\right\rangle \left\langle \omega,q,j\right|\;.\label{eq:rouRe-1}
\end{equation}
As the black hole evaporates, its mass decreases, and when $M$ becomes
very small, the effect of quantum gravity gradually begins to appear.
If the correction in the black hole entropy due to the quantum gravity
is taken into account, the area entropy for the black hole is modified
as \cite{key-4}
\end{widetext}

\begin{equation}
S_{H}=\frac{A_{H}}{4}+\alpha\ln\frac{A_{H}}{4}=\pi R_{H}^{2}+\alpha\ln\left(\pi R_{H}^{2}\right),\label{eq:S_QG}
\end{equation}
where $\alpha$ is a dimensionless parameter determined by the specific
quantum gravity model. The effect of quantum gravity causes Eq. (\ref{eq:rouRe-1})
to be rewritten as
\begin{widetext}
\begin{equation}
\rho_{\textrm{R}}=\sum_{\omega,q,j}\left[\frac{R_{H}\left(M-\omega,Q-q,J-j\right)}{R_{H}\left(M,Q,J\right)}\right]^{2\alpha}\textrm{e}^{-\left[\pi R_{H}^{2}\left(M,Q,J\right)-\pi R_{H}^{2}\left(M-\omega,Q-q,J-j\right)\right]}\left|\omega,q,j\right\rangle \left\langle \omega,q,j\right|.\;\label{eq:rouRe-quantum}
\end{equation}
For a given black hole, one can first get the horizon radius as the
function of its external qualities with the help of its metric. And
then by making use of Eqs. (\ref{eq:rouRe-1}) and (\ref{eq:rouRe-quantum}),
the radiation spectrum in the classical and quantum case are obtained,
respectively.

\textit{Schwarzschild black hole}.-As the simplest black hole, there
is only one hair for the Schwarzschild black hole, that is, the mass
$M$ . The radius of its event horizon is $R_{H}=2M$, which together
with Eq. (\ref{eq:rouRe-1}) give the radiation spectrum of the Schwarzschild
black hole as
\end{widetext}

\begin{equation}
\rho_{\textrm{R}}=\sum_{\omega}\textrm{e}^{-8\pi\omega\left(M-\omega/2\right)}\left|\omega\right\rangle \left\langle \omega\right|,\label{eq:dm-s-bh}
\end{equation}
where $\left|\omega\right\rangle $ is the eigenstate of the radiation.
It is seen from Eq. (\ref{eq:dm-s-bh}) that the probability

\begin{equation}
p\left(\omega,M\right)=\textrm{e}^{-8\pi\omega\left(M-\omega/2\right)}\label{eq:p}
\end{equation}
for the state $\left|\omega\right\rangle $ being in the distribution
is the same as the tunneling probability $\Gamma\left(\omega,M\right)=\exp\left[-8\pi\omega\left(M-\omega/2\right)\right],$
which is known as the Parikh-Wilczek spectrum \cite{key-PW}, as the
result of the WKB approximation through the perspective of quantum
tunneling. Note that we did not use the tunneling dynamics of the
radiation process at all, but we can obtain the non-thermal spectrum
Eq. (\ref{eq:dm-s-bh}). Here, we only require the area entropy of
black hole to be the function of energy. The simple derivation of
radiation shows that the entropy of black hole is the key quantity
to determine the radiation spectrum. 

\textit{Reissner-Nordstr�m black hole}.-For the Reissner\textendash Nordstr�m
black hole with mass $M$ and charge $Q$, the radius of its outer
event horizon obtained from the metrics reads $R_{H}=M+\sqrt{M^{2}-Q^{2}}$.
Substituting it into Eq. (\ref{eq:rouRe-1}), we obtain the non-thermal
spectrum of R-N black hole as
\begin{widetext}
\begin{equation}
\rho_{\textrm{R}}=\sum_{\omega,q}\exp\left\{ \pi\left[\left(M-\omega\right)+\sqrt{\left(M-\omega\right)^{2}-\left(Q-q\right)^{2}}\right]^{2}-\pi\left(M+\sqrt{M^{2}-Q^{2}}\right)^{2}\right\} \left|\omega,q\right\rangle \left\langle \omega,q\right|\;,\label{eq:dm-RN-bh}
\end{equation}
where $\left|\omega,q\right\rangle $ is the eigenstate of the radiation.
Equation (\ref{eq:dm-RN-bh}) is exactly consistent with the radiation
probability that obtained from the perspective of quantum tunneling
in Ref \cite{key-Zhaozheng}. Here, we show the density matrix for
Schwarzschild and R-N black holes without referring the dynamics of
the particle tunneling. Similar process for other black holes result
in exact matching between our results and that obtained from dynamics
analysis. This remind us that while considering the non-thermal effect
of the black hole radiation spectrum, there is an even deeper intrinsic
relationship between the quantum tunneling approach and our statistical
method. For example, in previous work, by using the radiation spectrum,
we successfully obtained the number of micro-states of the black holes
\cite{number of state}.
\end{widetext}

\textit{Information of black hole radiation}.-Now we go back to Eq.
(\ref{eq:densityRa}) to discuss the information carried by Hawking
radiation. The corresponding Von-Neumann entropy $S_{\textrm{R}}=-\textrm{Tr}\left(\rho_{\textrm{R}}\ln\rho_{\textrm{R}}\right)$
of R reads

\begin{equation}
S_{\textrm{R}}=S_{\textrm{B}}-\sum_{r}\textrm{e}^{-\Delta S_{\textrm{BB}'}\left(E_{r},E\right)}S_{\textrm{B}'}\left(E-E_{r}\right).\label{eq:SR1}
\end{equation}
Namely, $S_{\textrm{R}}=S_{\textrm{B}}-S\left(\textrm{B}'|\textrm{R}\right)$,
where $S\left(\textrm{B}'|\textrm{R}\right)=\sum_{r}\textrm{e}^{-\Delta S_{\textrm{BB}'}\left(E_{r},E\right)}S_{\textrm{B}'}\left(E-E_{r}\right)$
is the conditional entropy of $\textrm{B}'$. This shows that there
exists correlation between R and $\textrm{B}'$, as a result of
energy conservation. In the low energy limit, i.e., $E_{r}\ll E$,
the conditional entropy is approximated, by keeping the first order
of $E_{r}$, as $S\left(\textrm{B}'\textrm{|R}\right)\approx S_{\textrm{B}'}\left(E_{\textrm{B}'}\right).$
Here, $E_{\textrm{B}'}=E-E_{\textrm{R}}$ and $E_{\textrm{R}}=\sum_{r}\exp\left[-\Delta S_{\textrm{BB}'}\left(E_{r},E\right)\right]E_{r}$
are the the internal energy of $\textrm{B}'$ and R, respectively.
This indicates that the correlation between R and B could be ignored
when the energy of R is much smaller than that of B. At this time,
the radiation spectrum of the Schwarzschild black hole in Eq. (\ref{eq:dm-s-bh})
can be approximated as $\rho_{\textrm{R}}=\sum_{\omega}\exp\left(-8\pi M\omega\right)\left|\omega\right\rangle \left\langle \omega\right|$,
which is just the thermal radiation discovered by Hawking. So far
it becomes clear that the thermal spectrum of the black hole radiation
is due to the ignorance of the correlation information between the
black hole and its radiation. This is also the primary cause of the
black hole information loss. Analogously, it can be proved that the
conditional entropy for $\textrm{B}'$ of an arbitrary black hole
system is $S\left(\textrm{B}'\textrm{|R}\right)\approx S_{\textrm{B}'}\left(M',Q',J'\right),$
where $M'=M-\sum_{\omega}p\left(\omega\right)\omega$, $Q'=Q-\sum_{q}p\left(q\right)q$,
and $J'=J-\sum_{j}p\left(j\right)j$ are the average of the mass,
charge, and angular momentum of the black hole $\textrm{B}'$. The
condition for this approximation are $\sum_{\omega}p\left(\omega\right)\omega\ll M$,
$\sum_{q}p\left(q\right)q\ll Q$, and $\sum_{j}p\left(j\right)j\ll J$. 

The existence of correlation between sequences of emitted particles
have been proved via mutual information \cite{key-20,key-19,key-QYC}.
Yet, the proofs are based on case by case studies. We revisit this
proof with our general formalism. The mutual information for two emissions
$\omega_{1}$ and $\omega_{2}$ can be written, by definition \cite{key-mi},
as $I=\sum_{\omega_{1},\omega_{2}}p_{1,2}\ln\left[p_{1,2}/\left(p_{1}p_{2}\right)\right]$,
where $p_{1}$ ($p_{2}$) is the possibility for $\omega_{1}$ ($\omega_{2}$)
in the distribution, and $p_{1,2}$ is the joint probability of the
two emissions. In our case, these possibilities are given by Eq. (\ref{eq:p})
as $p_{1}=p\left(\omega_{1},M\right)$, $p_{2}=p\left(\omega_{2},M\right)$,
and $p_{1,2}=p\left(\omega_{1}+\omega_{2},M\right)$ . Moreover, it's
easily checked that $p\left(\omega_{1}+\omega_{2},M\right)=p\left(\omega_{1},M\right)p\left(\omega_{2},M-\omega_{1}\right)$,
with the help of which, by straightforward calculation, we obtain
the mutual information $I=8\pi\left\langle \omega_{1}\right\rangle \left\langle \omega_{2}\right\rangle ,$
where $\left\langle \omega_{i}\right\rangle =\sum\omega_{i}p\left(\omega_{i},M\right)$
is the average energy of $\omega_{i}$ ($i=1,2$). This shows that
the mutual information between the emissions is proportional to their
internal energy, and coincide with the result in Ref \cite{key-QYC},
where the authors proved that the sum of these mutual information
is just the total information of the initial black hole. From this
point of view, our result can naturally give the conclusion that the
black hole information is not lost, if the information correlation
between the radiation particles is taken into account. The derivation
of this conclusion, which needs to be emphasized here, is now dynamics-independent.

\textit{Conclusion}.-In summary, we straightforward derived the non-thermal
spectrum of the black hole radiation from the area law of entropy,
which only depends on a few external qualities of the black hole (known
as hairs). The derivation is based on the principle of canonical typicality,
without referring to the dynamics of quantum tunneling across the
horizon.We showed that there exists information correlation between
the black hole and its radiation, and thus the fact that the information
is not lost with the black hole's evaporating is clarified. Taking
into account the effect of quantum gravity, we achieved the modified
radiation spectrum through our universal protocol succinctly. Since
the general formalism we developed does not require the specific form
of the system and has been successfully applied to the black hole
to obtain its non thermal radiation spectrum, we therefore conjecture
it can also be applied to some non - relativistic systems, such as
ideal gas, black-body, etc., to obtain their radiation spectra. The
advantage of this approach is that there is no need to analyze the
particle dynamics in the radiation process, but it only refers to the
dependency of the entropy on the macro-conserved qualities of the
whole system.

\textit{Acknowledgments}.-Y. H. Ma thanks J. F. Chen and X. Wang for
helpful discussions. We thank Prof. D. Stojkovic for his kind letter
to let us know his works in black hole information paradox. This study
is supported by the National Basic Research Program of China (Grants
No. 2014CB921403 and No.2016YFA0301201), the NSFC (Grants No. 11421063
and No.11534002), and the NSAF (Grant No. U1530401). 

\textit{Note added}.- Recently, we applied the canonical typicality approach developed in this paper to obtain the radiation spectrum of Schwarzschild black hole in the case with dark energy [Nucl. Phys. B, 931 (2018) 418]. Some interesting results have been reported in this relevant work, showing the influences of dark energy on the non-thermal radiation of black holes

\end{document}